\documentclass[prl, twocolumn, showpacs]{revtex4}
\usepackage{amssymb}
\usepackage{amsmath}
\usepackage{epsfig}
\usepackage{bm}

\begin{document}

\title{ A magnetic resonance in high-frequency
viscosity of two-dimensional electrons }
\author{ P. S. Alekseev }
\affiliation{Ioffe  Institute,  194021  St.~Petersburg, Russia}

\begin{abstract}

 Two-dimensional (2D) electrons in high-quality
  nanostructures at low temperatures
can form a viscous fluid.
  We develop a theory of high-frequency
  magnetotransport in such fluid.
 The time dispersion of  viscosity
  should be taken into account at the frequencies
 about  and above the rate of electron-electron
  collisions. We show that the shear viscosity coefficients
 as    functions of magnetic field and frequency
  have the only resonance
   at the frequency equal to the doubled cyclotron frequency.
   We demonstrate that such resonance manifests itself in the plasmon damping.
    Apparently, the predicted resonance  is also responsible for the peaks
  and features  in photoresistance
  and photovoltage, recently observed  on the best-quality
  GaAs quantum wells.
  The last  fact should considered as  an important evidence of forming
   a viscous electron fluid  in such structures.


\end{abstract}

\maketitle

{\em 1. Introduction.} In solids with enough weak disorder
 a viscous  fluid consisting of
 phonons or conductive electrons can be formed
 at low temperatures.
 For realization of such hydrodynamic regime,
  the inter-particle  collisions conserving  momentum
 must be much more intensive than any other collisions
 which do not conserve  momentum.
      This idea was proposed many years ago
    for 3D materials with strong phonon-phonon and
      electron-phonon interactions \cite{Gurzhi,Gurevich}.
    The hydrodynamic regime of thermal transport in liquid helium and dielectrics
    was studied in sufficient detail \cite{Pitajevkii}.
 However, in those years
  there existed no enough pure solids where
  the hydrodynamic regime
  of charge transport could be realized.

 Recently,
 the crisp  fingerprints of forming a 2D viscous electron fluid
 and realization of hydrodynamic charge transport
  were discovered   in novel ultra-high quality
   materials: in 3D Weyl semimetals \cite{Weyl_sem_1,Weyl_sem_2}
  as well as in 2D nanostructures:
  graphene \cite{grahene,grahene_2} and GaAs quantum wells \cite{je_visc}.
   The most bright of such fingerprints
   is the giant negative magnetoresistance  effect,
   which was  discovered
   the best-quality GaAs quantum wells \cite{exps_neg_1,exps_neg_2,exps_neg_3,exps_neg_4}
  and on the  Weyl semimetal  WP$_2$ \cite{Weyl_sem_2}.
These experimental discoveries were accompanied by an extensive development
 of theory
   \cite{Gurzhi_Kalinenko_Kopeliovich,Spivak,Andreev,Mendoza_Herrmann_Succi,Tomadin_Vignale_Polini,je_visc,Levitov_et_al,Lucas,eta_xy,we_1,we_2,we_3,we_4,we_5}.

  The story of the giant negative magnetoresistance was very bright and non-trivial.
  Most of the conventional bulk transport theories predict either
absent or parabolic positive magnetoresistance.
 The most well-known bulk mechanism for negative magnetoresistance
 is the weak localization effect, which leads to a relatively moderate
 negative magnetoresistance
 in very weak fields for materials with enough strong disorder.
 The giant negative magnetoresistance effect, which is the decrease of resistance by 1-2
orders of magnitude in moderate magnetic fields,  seemed outstanding,
 surprising  and mysterious during 5 years after its discovering \cite{MI_conf}.

 In Ref.~\cite{je_visc}
  it was shown that the giant negative  magnetoresistance can be explained in
  details within the hydrodynamic model
  taking into account the dependence of  the electron viscosity
  coefficients on magnetic field and temperature.
 By this way, one can consider the best-quality GaAs quantum wells and similar
 materials as a novel type of solids where the gydridynamic regime of charge transport is realized and
  the  electron viscous fluid flows through
  a crystal lattice like water through a porous organic material.

 In this Letter we provide the second possible evidence
 that the hydrodynamic regime of charge transport
 in realized in the ultra-high mobility GaAs quantum wells.

 We develop a  theory of non-stationary hydrodynamic transport
of a 2D viscous electron fluid in magnetic field \cite{conf}.
We derive the Navier-Stocks equation for an ac viscous flow
  taking into account the time dispersion of viscosity.
 The obtained frequency-dependent viscosity
 coefficients have a resonance
 at the frequency equal to the doubled electron
 cyclotron frequency, $\omega = 2 \omega_c$,
 herewith the other harmonics of the cyclotron resonance
 are absent in the coefficients of the Navier-Stocks equation.
 So this resonance is a very special type of
 the high-order  cyclotron resonance
 related to the viscosity effect.
 It has the following physical nature.
  A viscous flow is controlled  by the diffusive-like
  transfer of the  electron momentum,
 which is accompanied  by  the presence of the viscous stress.
 The last
 varies in magnetic field as a product of
 two components of the electron velocity,
 thus it oscillates with the doubled cyclotron frequency.

  We demonstrate that the proposed {\em viscous resonance} manifests itself
  in the damping coefficient of magnetoplasmons
  and in absorbtion of an ac field by the electron fluid.
    We also argue
  that, apparently, the {\em  viscous resonance } is  responsible for the peaks
  and features
  at $\omega=2\omega_c$ in the photoresistance
  and the photovoltaic effects, recently observed  on the best-quality
  GaAs quantum wells \cite{exp_GaAs_ac_1,exp_GaAs_ac_2,exp_GaAs_ac_3}.
 So the viscous resonance together with the giant negative magnetoresistance
  evidence of forming a viscous electron fluid
  in moderate magnetic fields in the ultra-pure GaAs quantum wells.

{\em 2. Viscous flow in magnetic field.}
The momentum flux density tensor  (per one particle)
is defined as:
$ \Pi _{ij} (\mathbf{r},t)= m\langle v_i v_j \rangle $,
 where $m$ is the electron mass, $\mathbf{v} =(v_x,\,v_y)$
is the velocity of a single electron and the angular brackets
stand for averaging over the electron velocity distribution at a
given time $t$ and point $\mathbf{r} = (x,\,y)$.
 The hydrodynamic velocity in this notations is
  $ V _{i}(\mathbf{r},t) = \langle v_i  \rangle$.
The values $\mathbf{V}$ and  $\Pi _{ij} $ are proportional to the first and the second
  angular harmonics (by the electron velocity vector $\mathbf{v}$) of
 the electron distribution function
  $f(\mathbf{v};\mathbf{r},t)$ (see discussion in Refs.~\cite{LP_Kin,LL_Hydr,Steinberg, Aliev}).

If electrons weakly interact between themselves
and can be regarded as an almost ideal Fermi gas,
the hydrodynamic approach can be used
when the characteristic space scale,
$L$, of changing of $\mathbf{V}(\mathbf{r},t)$
 is far greater than, at least, one of the following lengths:
the electron mean  free path relative to electron-electron  collisions
$l_{ee}=v_F \tau_{ee}$; the electron cyclotron radius $R_c=v_F/\omega_c$;
the length of the path that free electron passes during
the characteristic period of changing of $\mathbf{V}(\mathbf{r},t)$,
$l_{\omega}=v_F/\omega$. Here $v_F$ is the Fermi velocity,
$\tau_{ee} $ is the electron-electron scattering time
 (its exact definition will be clarified below),
 $\omega_c$ is the cyclotron frequency, and $\omega$ is
 the characteristic frequency of a flow.
If one of these conditions is satisfied,
then inside the regions of the size
$ L $  the quasi-equilibrium
 distribution of electrons is formed and the flow can be described by
  the values $\mathbf{V}$ and  $\Pi _{ij} $.

The equation for the hydrodynamic velocity in zero magnetic flied is:
\begin{equation}
\label{dmV__dt}
m\frac{\partial V_i}{\partial t}=-\frac{\partial\Pi _{ij}}{\partial
x_j} -\frac{mV_i}{\tau}  + e E_i
\:.
\end{equation}
Here  $e$ is the electron charge,
 $\tau$ is the momentum relaxation time
 related to electron scattering on disorder or phonons \cite{ph_rates},
 and summation over repeating indices is assumed.
The momentum flux density tensor $ \Pi _{ij}$
 is equal to  $P\delta_{ij}-\sigma_{ij}$,
where $P$ is the pressure in the fluid,
 $ \delta_{ij} $ is the Kronecker delta symbol,
 and $\sigma_{ij}$ is the viscous stress tensor \cite{LL_Hydr}.

 For  slow flows which vary
 at a time scale much greater than the time  of relaxation
of the inequilibrium part of the
momentum flux density tensor, $\Pi_{ij}$ is given by \cite{LL_Hydr}:
\begin{equation}
\label{Pi_ij_0}
\Pi_{ij}^{(0)} = P\delta_{ij}
 -m  \left[ \eta
\left( V_{ij} - \frac{1}{2} \,\delta_{ij} V_{kk}  \right)
 + \frac{\zeta}{2 }   \,\delta_{ij} V_{kk}  \right]
\:,
\end{equation}
where
$
V_{ij} = \partial V_i/\partial x_j+
\partial V_j/\partial x_i
 $,
$ \eta$ are $\zeta$
are the shear   and the bulk  viscosity coefficients.
For the Fermi gas the last is relatively small: $ \zeta
\sim (T/\varepsilon_F)^2 \eta$ \cite{zeta_F_zidk},
 where $T$ is temperature and $\varepsilon_F$ is the Fermi energy.
 In this regard, we will neglect
 the bulk viscosity in further consideration.

Using Eqs.~(\ref{dmV__dt}) and (\ref{Pi_ij_0}),
 one obtains the Navier-Stocks equation
 in the linear by $\mathbf{V}$ regime:
\begin{equation}
\label{Navier-Stocks_B=0}
\frac{\partial \mathbf{V} }{\partial  t }=
 \frac{e}{m}  \, \mathbf{E}  -  \frac{\mathbf{V}}{\tau}
- \nabla P + \eta \, \Delta \mathbf{V}
 \:.
\end{equation}
In this study we take into
account the compressibility of the electron fluid. Thus
 one needs to supplement Eq. (\ref{Navier-Stocks_B=0})
 by the gas equation of state $P=P(n)$ (here $n$ is the electron density)
 and  by the continuity equation. The last
 in the linear regime has the form:
\begin{equation}
\label{continu_eq}
\frac{\partial  n  }{ \partial t } + n_0 \mathrm{div } \mathbf{V} = 0
\:,
\end{equation}
where  $n_0$ is the unperturbed electron density.

 The value given
by Eq. (\ref{Pi_ij_0}) is attained during the time $\tau_{ee}$, as
described by the Drude-like equation \cite{Kaufman}:
\begin{equation}
\label{Pi_ij_relaxation}
\frac{\partial\Pi_{ij}}{\partial t}
 = -\frac{1}{\tau_{ee}}\big(\Pi_{ij}-\Pi_{ij}^{(0)}\big)
\:.
\end{equation}
 Here $\tau_{ee}$ is the time  of relaxation of
 the second angular moment (by the electron velocity)
 of the electron distribution function.
 As a rule, it is related to electron-electron scattering.
 Hydrodynamic effects are significant
 for an  electron fluid in a solid if the  scattering on disorder or phonons
 is much less intensive than electron-electron scattering:
  $\tau_{ee} \ll \tau$ \cite{Gurzhi}. Formulas
  (\ref{dmV__dt}),  (\ref{Pi_ij_0}), and  (\ref{Pi_ij_relaxation})
are the whole system of equations describing nonstationary flows
 of a 2D viscous electron fluid in zero magnetic field.

 For a high-frequency flow with characteristic
 frequencies $\omega$
compared to  $1/\tau_{ee}$
 the relation between $\Pi_{ik} (\mathbf{r},t)$
 and $ V_{ik} ( \mathbf{r} ,t )$ is nonlocal by time.
 Owing to linearity of  the all equations,
we can decompose all the values  by the time harmonics
 proportional to $e^{-i\omega t }$. For each pair of harmonic
$\mathbf{V} (\mathbf{r},\omega) e^{-i\omega t } $ and
$\Pi_{ij}(\mathbf{r},\omega) e^{-i\omega t } $ we obtain from
 Eq.~(\ref{dmV__dt}),  (\ref{Pi_ij_0}), and (\ref{Pi_ij_relaxation})
the relations between the amplitudes
$\mathbf{V} (\mathbf{r},\omega)$ and $\Pi_{ij}(\mathbf{r},\omega)$.
 This relations  have the same form as
Eqs.~(\ref{Pi_ij_0}) and (\ref{Navier-Stocks_B=0}),
 but contain the amplitude $\mathbf{E} (\mathbf{r},\omega)$
  of the electric field harmonic
 instead of $\mathbf{E} (\mathbf{r},t )$ and
the frequency-dependent viscosity coefficient,
$ \eta (\omega) = \eta/(1-i\omega \tau_{ee}),
 $ instead of $\eta$.

In the presence of magnetic field additional terms will appear in
the equations for $\partial V_i / \partial t $ and $\partial
\Pi_{ij} / \partial t $, since now the quantities  $\langle v_i
\rangle$  and $\langle v_i v_j \rangle$ will change in time not only
due to collisions and the electric field force, but also due to the
magnetic field force. The last force for  each electron is
 $ (eB/c) \epsilon_{ikz} v_k $,  where $\epsilon_{lik}$ is the unit
antisymmetric tensor and $z$ is the direction of the magnetic field
$\mathbf{B}$, which is perpendicular to the 2D electron layer.
 For the averaged products of the velocity components
 in the presence of only the magnetic field  $\mathbf{B}$ we have:
\begin{equation}
\label{Pi_ij_Newton}
\begin{array}{l}
\displaystyle
\frac{\partial \langle v_i  \rangle}{\partial t}  =
\omega_c \epsilon_{ikz} \langle v_k  \rangle
\:,
\\
\\
\displaystyle
\frac{\partial \langle v_i v_j \rangle}{\partial t} =
\omega_c  \big( \epsilon_{ikz}\langle v_k v_j \rangle +\epsilon_{jkz}\langle v_i v_k\rangle\big)
\:.
\end{array}
\end{equation}
The terms (\ref{Pi_ij_Newton}) should be added to the right-hand
side of Eqs.~(\ref{dmV__dt}) and (\ref{Pi_ij_relaxation}) \cite{E_ll_B}:
\begin{equation}
\label{Pi_ij_relaxation_B}
\begin{array}{l}
\displaystyle
m\frac{\partial V_i}{\partial t}= -\frac{mV_i}{\tau}
-\frac{\partial\Pi _{ij}}{\partial x_j}
+ eE_i + \omega_c \epsilon_{ikz} V_k
\:,
\\
\\
\displaystyle
\frac{\partial\Pi_{ij}}{\partial t}
 = -\frac{\Pi_{ij} -  \Pi_{ij}^{(0)}}{\tau_{ee}}
+  \omega_c  \big( \epsilon_{ikz}\Pi_{kj} +\epsilon_{jkz} \Pi_{ik} \big)
\:.
\end{array}
\end{equation}

As in the case of zero magnetic field, we,  first, consider
the case of slow flows when
the characteristic frequencies of $\mathbf{V}(\mathbf{r},t)$
 are small in comparison with $\omega_c$ and $1/\tau_{ee}$.
Putting $\partial \Pi_{ij}/ \partial t=0$, we find from
Eqs.~(\ref{Pi_ij_0}) and (\ref{Pi_ij_relaxation_B})
the values  $\Pi_{ij}$ as a linear combination of the values
 $\Pi_{ij} ^{(0)}$ and, thus, of $P$ and $V_{ij}$:
\begin{equation}
\label{def_tensor_eta}
\begin{array}{c}
\displaystyle
\Pi_{ij} = P \delta_{ij}   -  \sigma_{ij}
\:,
\\
\\
\displaystyle
 \sigma_{ij} =
 m \, \Big[
\eta_{xx}
\Big( V_{ij} - \frac{1}{2} \delta _{ij} V_{kk} \Big)
 + \frac{\eta_{xy}}{2}  \epsilon_{ ikz} V_{kj}
 \Big]
  \:
  \end{array}
\end{equation}
where $\eta_{xx}$ and $\eta_{xy}$ are the stationary
 shear viscosity  coefficients
of 2D electron fluid in magnetic field (see Ref.~\cite{je_visc}
 and Eq.~(\ref{eta zeta ot omega}) at $\omega = 0 $).

With the help of Eqs.~(\ref{Pi_ij_relaxation_B}) and (\ref{def_tensor_eta}),
we arrive to the Navier-Stocks equation
 of the compressible 2D electron fluid in magnetic field at low frequencies,
  which differs from Eq.~(\ref{Navier-Stocks_B=0})
  by the change of $\eta $ on $\eta_{xx}$
  and the appearance the two magnetic terms $
   \omega_c \, [ \mathbf{V} \times \mathbf{e}_z] $
  and $
  \eta_{xy} \, [\Delta \mathbf{V} \times \mathbf{e}_z ]
  $ \cite{je_visc}.

 \begin{figure}[t!]
\centerline{\includegraphics[width=0.8\linewidth]{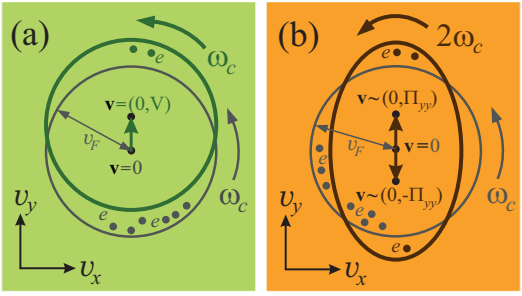}}
\caption{ Schematic representation of the two
 distributions $f(\mathbf{v})$ of 2D electrons by
 their velocities $\mathbf{v}=(v_x,v_y)$.
  The equilibrium  Fermi distributions $f_F(\mathbf{v})$ are shown by grey
  on both panels (a) and (b).
 The quasi-equilibrium distribution,
 $f_{\mathbf{V}}(\mathbf{v})=f_F(\mathbf{v}-\mathbf{V})$, with
 the mean  hydrodynamic velocity $\mathbf{V}=(0,V) $
 is shown by green at panel (a).
 In magnetic field such distribution function,
 together each individual electron, rotates
 with the frequency $\omega_c$.
 The nonequilibrium distribution, $f_{\Pi}(\mathbf{v}) = f_F(\mathbf{v}) + f_2(\mathbf{v}) $,
 with zero mean velocity and
 the second harmonic $f_2(\mathbf{v})$ corresponding
 to a non-zero  component $\Pi_{yy}$ of the  momentum flux density tensor $\Pi_{ij}$ is shown
 by brown on panel (b). The rotation of individual electrons
 with the frequency $\omega_c$ leads to
 the rotation of such distribution function $f_{\Pi}$
 with the frequency $2 \omega_c$.
 }
\end{figure}

Second, we consider the case of a high-frequency flow when
the characteristic frequencies $\omega$ are
compared to $\omega_c$ and $1/\tau_{ee}$.
 As in the case of zero magnetic field,
   we decompose the values  $\mathbf{V}(\mathbf{r},t)$ and
$\Pi_{ij}(\mathbf{r},t)$ by the time harmonics
proportional to $e^{-i\omega t }$. As a final result,
 we arrive to the Navier-Stocks equation  for the amplitude
$\mathbf{V} (\mathbf{r},\omega) $ of each velocity harmonic:
\begin{equation}
\label{Navier_Stocks_B_with_time_disp}
 \begin{array}{c}
 \displaystyle
-i\omega  \mathbf{V}
=
 \frac{e}{m}\mathbf{E}(\mathbf{r},\omega)
  + \omega_c \left[\mathbf{V} \times \mathbf{e}_z \right]
    -\frac{\mathbf{V}}{\tau}
 -\frac{\nabla P }{m} +
\\
\\
\displaystyle
+ \eta_{xx} (\omega) \, \Delta \mathbf{V}+  \eta _{xy} (\omega)
  \left[  \Delta \mathbf{V} \times \mathbf{e}_z \right]
  \end{array}
  \:,
\end{equation}
where   the viscosity coefficients
 depend on magnetic field and frequency:
\begin{equation}
\label{eta zeta ot omega}
\begin{array}{c}
 \displaystyle
\eta_{xx} (\omega) =\eta \frac{1- i\omega\tau_{ee}}
 {1+(-\omega^2+ 4\omega_c^2 )\tau_{ee}^2 - 2i\omega\tau_{ee}}
  \:,
\\
 \\
\displaystyle
\eta_{xy} (\omega) =\eta \frac{2\omega_c\tau_{ee}}
 {1+(-\omega^2+ 4\omega_c^2 )\tau_{ee}^2 - 2i\omega\tau_{ee}}
 \:.
\end{array}
\end{equation}

It is seen that at  $\omega_c \gg 1/\tau_{ee}$ the viscosity coefficients
 $\eta_{xx} (\omega) $ and $\eta_{xy} (\omega) $
 exhibit a resonance at  $\omega=2 \omega_c$.
 Indeed, the own
 frequency of rotation of the value $\Pi_{ij} = m \langle v_i v_j \rangle $
 is the doubled cyclotron frequency $ 2\omega_c $ (see Fig.~1).
 Thus when the frequency $\omega$ of
 variation of a flow is close to the internal frequency $2 \omega_c$,
  the resonance occurs.
 It is not just a second harmonic of the one-particle cyclotron resonance, as it is related not to
 motion of individual electrons, but to the motion of the momentum flux of the electron
 ensemble (see Fig.~1).
 Such resonance is the special type of the high-order cyclotron resonance of
 collective electron motion related to the viscosity effect in magnetic field
 and so it can be called {\em the viscous resonance}.

If the interaction between   2D electrons is strong,
 they must be treated as a Fermi liquid.
 The Navier-Stocks equation
  (\ref{Navier_Stocks_B_with_time_disp}), apparently,
   will describe flows of the fluid consisting
   of the quasiparticles of the Fermi liquid. The
coefficients $\eta$ and  $\zeta$ will contain the Landau parameters
 describing the interaction between quasiparticles.
 A preliminary analysis, following to Ref.~\cite{zeta_F_zidk}, shows that
  the conditions of applicability of
   the theory will expand significantly. In particular,
   the equations (\ref{Navier_Stocks_B_with_time_disp})
   and (\ref{eta zeta ot omega}) will be applicable even at short wavelengths
   and high frequencies, $L\sim l_{\omega}$.

{\em 3. Plasmon damping.} The time dispersion of viscosity
can manifest itself in damping of the magnetoplasmons.
 Below we calculate the magnetoplasmon damping coefficient
 related to viscosity using
 the equations (\ref{continu_eq}), (\ref{Navier_Stocks_B_with_time_disp}),
 and (\ref{eta zeta ot omega}).
 Herewith, we will not consider the retardation effects
 which can be important in the region of small wavevectors in some structures
 (see, for example, Ref.~\cite{Falko_Khmelnitski,Volkov_Zabolotnykh}).

 For the case of waves in the absence of external ac fields,
  the electric field $\mathbf{E}( \mathbf{r},\omega )$
  in Eq.~(\ref{Navier_Stocks_B_with_time_disp}) is
induced by the perturbation of
the 2D electron density $\delta n = n - n_0$.
When we can neglect the retardation effects, we just have
 $ \mathbf{E} = - \nabla \delta \varphi $,
 where $\delta \varphi $ is related to $\delta n$
  by the electrostatic equations.
 For the structures with a metallic gate located at
 the distance $d$ from the 2D layer we have:
$
\delta \varphi  = ( 4 \pi e d   /\kappa ) \, \delta n
$,
where $\kappa$ is the background dielectric constant.
For the structures without  a  gate the relation
 between $\delta \varphi(\mathbf{r},t)$ and $\delta n(\mathbf{r},t)$
 is given just by the Coulomb law with the charge density
 $\varrho(\mathbf{r},z) =  e \,  \delta n (\mathbf{r}) \, \delta(z) $, where
 $\delta(z)$ is the Delta-function depicting the position of the 2D layer.

We solve the together the equations (\ref{continu_eq}),
(\ref{Navier_Stocks_B_with_time_disp}),
and the electrostatic equation assuming that
$
\delta n (\mathbf{r},t)
 \,,\;
 \delta \varphi (\mathbf{r},t)
 \,,\;
\mathbf{V} (\mathbf{r},t) \sim  e^{-i \omega t + \mathbf{q} \cdot \mathbf{r}}
$.
The ratio of the terms $-\nabla P /m $ and $e\mathbf{E} /m $
 in Eq.~(\ref{Navier_Stocks_B_with_time_disp}) is estimated
as $a_B /d $ for the structures with a gate
and as $a_B q    $ for the ungated structures,
 where $a_B$ is the Bohr radius.
Both these values must be much smaller than
 unity when the 2D electrostatic equations
 are applicable. Neglecting  the terms describing  the relaxation processes,
we obtain from Eqs.~(\ref{continu_eq}) and  (\ref{Navier_Stocks_B_with_time_disp})
 the usual formula
 for the dispersion law of magnetoplasmons.
 For the gated structures it is:
\begin{equation}
\label{omega_0}
 \omega_{0,q} =  \sqrt{\omega_c^2 + s^2 q ^2 }
\end{equation}
where  $ s = \sqrt{ 4 \pi e^2 n_0 d  / m \kappa} $.
The second term under the root in Eq.~(\ref{omega_0}), $s^2q^2$,
 is the squared plasmon frequency in the absence of magnetic field.
For the ungated structure
it changes on  $ 2 \pi e^2 n_0   q/ m \kappa$.

\begin{figure}[t!]
\centerline{\includegraphics[width=0.8\linewidth]{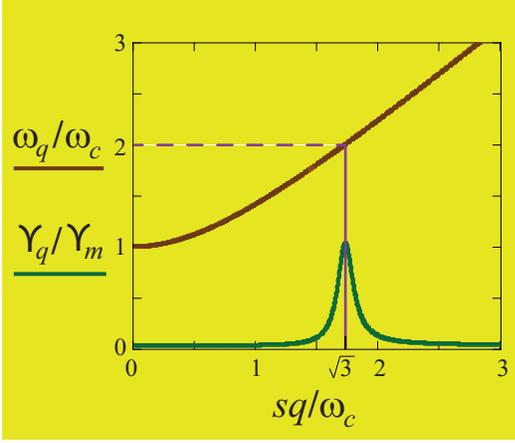}}
\caption{ The magnetoplasmon dispersion law  $\omega_{0,q}$
 and the damping coefficient $\Upsilon_q$
 calculated by Eqs.~(\ref{omega_0}) and (\ref{res})
  for a gated structure. The maximum value of $\Upsilon_q$,
  $\Upsilon_m = 5/(8\tau) + 27 \eta \omega_c ^2 /(8s^2)$ is attained at
  the wavevector $q_m = \sqrt{3} \omega_c/s$ corresponding to
  the equality $\omega_{0,q} =2\omega_c $.  }
\end{figure}

The viscosity terms and the terms describing
 scattering on disorder  leads to a small correction
 to the magnetoplasmon dispersion (\ref{omega_0})
  as well as to arising of  a finite damping:
$
\omega _{q}
 = \omega_{0,q} + \Delta \omega_q
  - i \,
   \Upsilon_q
   $.
The damping coefficient $\Upsilon_q $ takes the form:
\begin{equation}
\label{Ips_gen}
\begin{array}{c}
\displaystyle
\Upsilon_q
 =
   \frac{\omega_c ^2 + \omega_{0,q} ^2 }{2 \omega _{0,q}^2 }
 \left[ \frac{1}{\tau} +
 \mathrm{Re } \, \eta _{xx}  q^2
 \right]
   +
   \frac{\omega_c }{\omega_{0,q} }
    \:\mathrm{Im } \, \eta _{xy}  q^2
    ,
 \end{array}
\end{equation}
Here the viscosity coefficients
$\eta _{xx} (\omega) $
and
$ \eta _{xy} (\omega) $
are taken at
$\omega =  \omega _{0,q} $.

At high frequencies and high magnetic fields, $\omega, \omega_c \gg 1/\tau_{ee}$,
 we obtain from   Eqs.~(\ref{eta zeta ot omega}) and (\ref{Ips_gen}):
\begin{equation}
\label{res_dlinn}
\begin{array}{c}
\displaystyle
\Upsilon_q
 = \frac{1}{\tau} \frac{w^2 + 1}{2w^2}
+   \frac{ \eta q^2 }{ w^2 }
 \frac{ w^4  + 13 w^ 2 + 4 }{  4 w^2 + \beta^2 ( w^2 - 4)^2  }
 \:,
 \end{array}
\end{equation}
where $\beta = \omega _c \tau_{ee} \gg 1  $
and $w = w(q)= \omega_{0,q} / \omega_c $.
Near the resonance of the shear viscosity coefficients, when $w \approx 2$,
 the value $\Upsilon_q$ takes the form:
\begin{equation}
\label{res}
\Upsilon_q =
 \frac{5}{ 8\tau }
 +
 \frac{ 9 \eta q^2}{ 8( 1 +\varepsilon^2 \beta^2 ) }
 \:.
\end{equation}
where $\varepsilon = \varepsilon(q) = w (q)-2 $, $\varepsilon \ll 1$.
 In high-quality structures at low temperature the inequality
 $1/\tau  \lesssim \eta q^2/ \beta^2 $
 can take place in certain intervals of wavevectors
 and magnetic fields. Provided this condition,
 the damping coefficient  $\Upsilon_q$
 in the resonance is  greater than outside the resonance
 in $\beta^2  \gg 1 $ times [see Eq.~(\ref{res}) and  Fig.~2].

{\em 4. Discussion and  conclusion.}
 In the case when a viscous flow of a 2D electron fluid is induced by an external
  ac electric  field $\mathbf{E}_{ex} ( \mathbf{r} ,t)\sim e^{-i\omega t}$,
  the viscosity effect, together with electron scattering on disorder,
 determines the absorbtion of energy from the external field.
 The linear response of a  2D fluid on $\mathbf{E}_{ex} ( \mathbf{r} ,t)$
 should be calculated from Eqs.~(\ref{continu_eq})
 and  (\ref{Navier_Stocks_B_with_time_disp}).
 The resulting absorbtion coefficient will reflect the resonance dependence (\ref{res})
 of the magnetoplasmon damping, if the character plasmon wavelength $2 \pi /q_m$
 at the resonance frequency $\omega=2\omega_c$ is smaller that the sample width $W$.

 It is possible that  the viscous resonance is  responsible
 also for the strong peak and features
 observed at $\omega = 2 \omega_c$
  in the photoresistance  \cite{exp_GaAs_ac_1,exp_GaAs_ac_2}
  and the photovoltaic effects \cite{exp_GaAs_ac_3}
 in the high-mobility GaAs quantum wells.
  Indeed, it was stressed  in Ref.~\cite{exp_GaAs_ac_1}
  that the strong peak in photovoltage
  and the very well pronounced giant negative magnetoresistance,
 explained in Ref.~\cite{je_visc}
 as a manifestation of forming of a viscous flow,
  are observed in the {\em same best-quality} GaAs structures.
   If a 2D electrons
   in such structures form a viscous fluid, than any response of the structure
   on ac field
  (absorbtion, photovoltage, photoresistance)
  must inevitably have  peculiarities
  at the frequency of the viscous resonance.

 To construct the theories of the photoresistance
  and the photovoltaic effects, one should
 supplement  the hydrodynamic equation (\ref{Navier_Stocks_B_with_time_disp})
  by the nonlinear terms   following to Refs.~\cite{Lifshits_Dyakonov,Beltukov_Dyakonov}.
   The peak and features at $\omega = 2 \omega_c$ in  photovoltage and  photoresistance
 was observed in Refs.~\cite{exp_GaAs_ac_1,exp_GaAs_ac_2,exp_GaAs_ac_3} at rather high magnetic fields
 when the inequality $R_c \ll W$ is fulfilled. A preliminary analysis shows that this justifies
 the applicability of the Fermi-gas model for the
 description of hydrodynamics  near the viscous resonance.
  However, the Fermi-gas model  outside the resonance,
   in particular, in small magnetic fields,
  seems to be irrelevant.
  Justification of the realization of hydrodynamics outside the resonance,
  possibly, within the Fermi-liquid model, requires further study.

To conclude, we predict the viscous resonance at $\omega=2\omega_c$ related
to motion of the viscous stress tensor in magnetic field. This resonance manifest itself
in the dependence of the damping of magnetoplasmons on their wavevectror and, probably, in
the photoresistance and the photovoltaic effects.

The author wishes to thank Professor M.~I.~Dyakonov,
under whose guidance this research was undertaken; for the
discussions, advice, and support during the course of the work; and for
his participation in writing the text of the Letter.
 The author  also thanks A.~P.~Dmitriev and I.~V.~Gorniy for valuable discussions;
 D.~G.~Polyakov for attracting his attention to  Ref.~\cite{Kaufman};
  A.~P.~Alekseeva, E.~G.~Alekseeva, I.~P.~Alekseeva,
  N.~S.~Averkiev, A.~I.~Chugunov, M.~M.~Glazov, I.~V.~Krainov, A.~N.~Poddubny,
 P.~S.~Shternin, D.~S.~Svinkin, and V.~A.~Volkov  for advice and support.
 The part of this work devoted to the time dispersion of viscosity in magnetic field  (Section 2)
 was supported by
 the Russian Science Foundation (Grant No.  17-12-01182);
 the part of this work devoted to plasmon damping due to viscosity
 (Section 3) was supported by the grant of the Basis Foundation.


\begin{thebibliography}{}
\bibitem{E_ll_B}
In the second of Eq.~(\ref{Pi_ij_relaxation_B})
  we assume that the electric
field force is far smaller than the magnetic field force,
 as it usually takes place
in the magnetotransport experiments.
\end{thebibliography}
\end{document}